\documentclass[pamm,a4paper,fleqn]{w-art}
\usepackage{times,cite,w-thm}
\usepackage[T1]{fontenc}
\usepackage[utf8]{inputenc}
\usepackage{amsfonts}
\usepackage{mathabx}
\usepackage[colorlinks=true, allcolors=blue]{hyperref}
\usepackage{graphicx}
\usepackage{bbm}
\begin{document}
%% \def\leftmark{Session title}
%%
%%    The information for the title page will be placed between
%%    \begin{document} and \maketitle. The order of most entries
%%    is determined by the class file and cannot be changed by
%%    rearranging them. The maketitle command follows after the
%%    abstract.
%%
%%    The following commands will be updated by the publisher:
%%
%%    \renewcommand{\copyrightyear}{2016}
%%    \DOIsuffix{pamm.20161zzzz}
%%    \Volume{16} 
%%    \Year{2016} 
%%    \pagespan{1}{}
%%
%%    The short title is optional:

\TitleLanguage[EN]
\title[Evaluation of the elastic field in phase-field crystal models]{Evaluation of the elastic field in phase-field crystal simulations}

%% Please delete not needed author entries.
%% Information for the first author.
\author{\firstname{Maik} \lastname{Punke}\inst{1,}%
\footnote{Corresponding author: e-mail \ElectronicMail{maik.punke@tu-dresden.de}, 
     phone +49\,351\,463-41202,
     fax +49\,351\,463-37096}}

\author{\firstname{Vidar} \lastname{Skogvoll}\inst{2}}%
\author{\firstname{Marco} \lastname{Salvalaglio}\inst{1,3}%
    % \footnote{Second author footnote.}
    }
    
\address[\inst{1}]{\CountryCode[DE]Institute of Scientific Computing, TU Dresden, 01062 Dresden, Germany}
\address[\inst{2}]{\CountryCode[NO]PoreLab, Njord Centre, Department of Physics, University of Oslo, P. O. Box 1048, 0316 Oslo, Norway}
\address[\inst{3}]{\CountryCode[DE]Dresden Center for Computational Materials Science, TU Dresden, 01062 Dresden, Germany}
%%

%%
%%    \dedicatory{This is a dedicatory.}
%%
%%    Abstract is required.
\AbstractLanguage[EN]
\begin{abstract}
The phase-field crystal model (PFC) describes crystal structures at diffusive timescales through a periodic order parameter representing the atomic density. One of its main features is that it naturally incorporates elastic and plastic deformation. To correctly interpret numerical simulation results or devise extensions related to the elasticity description, it is important to have direct access to the elastic field. In this work, we discuss its evaluation in classical PFC models based on the Swift-Hohenberg energy functional. We consider approaches where the stress field can be derived from the microscopic density field (i.e., the order parameter), and a simple novel numerical routine is proposed. By numerical simulations, we demonstrate that it overcomes some limitations of currently used methods. Moreover, we shed light on the elasticity description conveyed by classical PFC models, characterizing a residual stress effect present at equilibrium. We show explicitly and discuss the evaluation of the elastic fields in prototypical representative cases involving an elastic inclusion, a grain boundary, and dislocations.
\end{abstract}

\maketitle

\section{Introduction}
\label{sec:model}
The so-called phase-field crystal (PFC) model \cite{Elder2002,Elder2004,Emmerich2012}
emerged as a prominent framework for describing crystal structures at large (diffusive) timescales through a continuous, periodic order parameter representing the atomic density. In its classical formulation, it builds on a Swift-Hohenberg-like free energy functional \cite{Elder2002,Elder2004,Emmerich2012} of the form
    \begin{equation}
    \begin{split}
    F_\psi\left[\psi\right] = \int_\Omega f_\psi[\psi]  \rm{d}\boldsymbol{r}, \qquad \qquad
f_\psi[\psi] = \dfrac{\lambda-\kappa}{2}\psi^2 -  \dfrac{\delta}{6}\psi^3 + \dfrac{\psi^4}{12}+ \dfrac{\kappa}{2}\psi \mathcal{L}^2 \psi,
\end{split}
\label{eq:F1}
    \end{equation}
with $f_\psi[\psi]$ the free energy density.
The scalar order parameter $\psi\equiv \psi(\boldsymbol{r},t)$ is related to the atomic number density, and $\Omega \in \mathbb{R}^n$ ($n=2$ in this work, without loss of generality). A set of parameters $\lambda,\,\kappa,\, \delta \geq 0$ characterizes the phase space and material properties. $\lambda$ plays the role of the quenching depth, while $\kappa$ is a constant that scales the elastic moduli expressed here in dimensionless units \cite{Punke_2022}. $\mathcal{L}$ is a differential operator of the form $\mathcal{L} = \mathcal{L}_1 \equiv (\beta(\boldsymbol{r})^2+\nabla^2)$ for a 2D triangular (or 3D body-centered cubic) crystal and $\mathcal{L}=\mathcal L_1 \mathcal{L}_2 \equiv (\beta(\boldsymbol{r})^2+\nabla^2)(2\beta(\boldsymbol{r})^2+\nabla^2)$ for a 2D square symmetric crystal. A spatially dependent length scale \cite{counterexamples,Punke_2022} of $2\pi/q(\boldsymbol{r})$ with $q(\boldsymbol{r})= \beta(\boldsymbol{r})q_0$ and $q_0 = 1$ is considered below. 
This approach describes a spatially dependent lattice parameter as it occurs, e.g., in heterostructures and in the presence of temperature gradients. 
Examples involving 2D square symmetry, however, will be discussed for the more classical model where $\beta(\boldsymbol{r})\equiv 1$, i.e., $\mathcal{L}_1 \mathcal{L}_2 \equiv (1+\nabla^2)(2+\nabla^2)$.  
In classical PFC methodology, the free energy is constructed to favor a limited Fourier expansion of the density field in equilibrium, which is usually truncated at the smallest reciprocal lattice vectors for the 2D triangular and 3D bcc model (the \textit{one-mode approximation}), and at the second smallest reciprocal lattice vectors for the 2D square model (the \textit{two-mode approximation}). 
Together with appropriate boundary and initial conditions, the dynamical equation for $\psi$, is given by a mass-conservative ($\mathrm{H}^{-1}$) gradient flow of $F_\psi$. The time evolution reads
\begin{equation}
    \begin{split}
\partial _t \psi = \nabla^2 \mu_\psi,\qquad \qquad
\mu_\psi =\dfrac{\delta F_\psi\left[\psi\right]}{\delta \psi}= \left(\lambda-\kappa+\mathcal{L}^2\right)\psi-\dfrac{\delta}{2}\psi^2+\dfrac{1}{3}\psi^3.
    \end{split}
    \label{eq:pfc}
    \end{equation}
This approach describes several phenomena, including crystal growth, phase transformations, grain boundaries, and dislocations \cite{Elder2002,Elder2004,Emmerich2012}. However, in its basic formulation, lattice distortions evolve at diffusive dynamics too, which is often an unphysical limit. Several extensions have been proposed to tackle more accurate modeling of elastic relaxation \cite{Stefanovic2006,heinonen2014,heinonen2016,skaugen2018,skogvoll2021,skogvollhpfc,Burns_2022}.

We consider the versatile hydrodynamic phase-field crystal approach (hPFC), recently proposed to account for the relaxation of elastic excitations through the coupling of the classical governing equation with the dynamics of a macroscopic velocity field $\boldsymbol{v}$ \cite{skogvollhpfc}. In brief, this model builds on the linear response theory for the relaxation to equilibrium with a free energy
    \begin{equation}
    \begin{split}
F[\psi,\boldsymbol{v}] = F_\psi[\psi] + \int_\Omega \dfrac{\psi}{2}\left|\boldsymbol{v}\right|^2 \rm{d}\boldsymbol{r}.
    \end{split}
    \label{eq:hpfc_energy}
    \end{equation}
Governing equations are obtained under generic conservation laws for the evolution of mass and momentum. The resulting system of equations under the assumption of constant average density (namely for bulk systems) reads
%a macroscopic velocity field $\boldsymbol{v}\in \mathbb{R}^2$:
%The fast dynamics of $\boldsymbol{v}$ is coupled with the diffusive evolution of the PFC, which leads to a system of equations
    \begin{equation}
    \begin{split}
\partial _t \psi &= \nabla^2 \mu_\psi -\boldsymbol{v}\cdot \nabla \psi,\\
\rho \partial_t \boldsymbol{v} &=\langle \mu_\psi \nabla \psi - \nabla f_\psi\rangle + \Gamma \nabla^2 \boldsymbol{v},
    \end{split}
    \label{eq:hpfc}
    \end{equation}
 with parameters $\rho$ and $\Gamma$ controlling the speed of elastic relaxation. The operator $\langle\cdot \rangle$ stands for a smoothing kernel
   \begin{equation}
     \langle f \rangle = \mathbb{F}^{-1}(\rm{e}^{-2 \pi^2 \boldsymbol{k}^2}\mathbb{F}( f)),
    \label{eq:smooth}
    \end{equation}
    with $\mathbb{F}$ the spatial Fourier transform and $\mathbb{F}^{-1}$ the inverse spatial Fourier transform of the periodic function $f$ with spatial Fourier vector $\boldsymbol{k} =(k_1,k_2)$. The smoothing kernel includes a coarse-graining procedure over one unit cell $\rm{UC}=[0,a_1]\times [0,a_2]$ with $a_1 = p$, $a_2 = \sqrt{3}p/2$, $p=4\pi\sqrt{3}$ for the triangular symmetry and $a_1=a_2=p=2\pi$ for the square symmetry.
    
In this article, we cover practical and fundamental aspects of evaluating the elastic field in the classical PFC model.
We show how to extract the elastic field from the density field and that the hPFC method relaxes elastic waves efficiently so that these stress fields satisfy the mechanical equilibrium condition. These fields are shown, for instance, to match those predicted by continuum elasticity for the Eshelby inclusion problem --- up to a residual isotropic pressure which may be present in the considered settings for PFC simulations. 
A residual pressure results from the density field being slightly compressed (or stretched) at equilibrium in the periodic simulation domain commonly considered for PFC simulations.
We characterize this effect in detail and compare the stress field produced by the density field obtained directly from PFC simulations with a modified one where higher modes in the Fourier spectrum have been removed. 
The stress fields produced by this \textit{filtered} density quantitatively closely matches those predicted by continuum elasticity.
Other approximations, which specify a small set of reference Fourier modes of a relaxed crystal and evaluate the deformation through their complex amplitudes via demodulation are considered in literature~\cite{Punke_2022,skogvoll2021}. 
However, these approaches lack generality when dealing with bicrystals and polycrystals, which inherently need more than one reference crystal structure. Moreover, they lead to a significant averaging of small-scale features of the stress fields. The filtered density field mentioned above is shown to overcome the limitations for bicrystals and dynamical configurations with dislocations featuring non-trivial elastic fields.

\section{Evaluation of elastic fields}
Lattice distortions in terms of the deviation of $\psi$ from an undistorted configuration can be described through the classical notion of displacement field $\boldsymbol{u}$. This field usually varies over a length scale significantly larger than the lattice spacing. Therefore, coarse-graining procedures as in Eq.~\eqref{eq:smooth} are usually considered in this context~\cite{skaugen2018,salvalaglio2019npj,salvalaglio2020,skogvoll2021} (and indeed underly the derivation of Eq.~\eqref{eq:hpfc} \cite{skogvollhpfc}). 
One may obtain the mechanical stress tensor $\sigma_{ij}\equiv\sigma_{ij}(\boldsymbol{r},t)$ from the density field as \cite{skogvoll2021}:
\begin{equation}
    \begin{split}
\sigma_{ij} &=\bigg \langle\dfrac{\delta F}{\delta \partial_i u_j}\bigg \rangle,
    \end{split}
    \label{eq:stress}
    \end{equation}
with $\delta F = F[\psi(\boldsymbol{r}-\boldsymbol{u},t), \boldsymbol{v}(\boldsymbol{r},t)]- F[\psi(\boldsymbol{r},t), \boldsymbol{v}(\boldsymbol{r},t)]$.
%and implying Einstein's summation convention. 
While we explicitly discuss the stress field, the strain field can be obtained via analogous procedures and constitutive relations.
Definition \eqref{eq:stress} translates for a triangular lattice with spatially dependent lattice constant $\beta$ into:
\begin{equation}
    \begin{split}
\sigma_{ij} &=\kappa \left( \left(2\beta^2 \psi_{,j} + \psi_{,kki} + \psi \beta^2_{,i}\right)\psi_{,j} -\psi_{,kk}\psi_{,ij}\right),
    \end{split}
    \label{eq:stress_tri}
    \end{equation}
where the Einstein summation convention is assumed.
For the square symmetry with lattice constant $\beta \equiv 1$, Eq.~\eqref{eq:stress} becomes:
  \begin{equation}
\sigma_{ij} =-2\kappa \left(2\psi+3\psi_{,kk} + \psi_{,kkpp} \right) \left(3\psi_{,ij}+2\psi_{,rrqqij}\right).
    \label{eq:stress_square}
    \end{equation}
 
For any finite quenching depth, the equilibrium lattice unit length will slightly differ from the prescribed value $p$ given by $\beta(\boldsymbol r)$. 
Thus, the PFC will experience a slight residual isotropic stress if initiated in a simulation domain commensurate with an integer multiple of $p$.
As we will show, the resulting compression (or stretching) manifests through higher-order modes in the Fourier spectrum of the density field and disappears when filtering out these higher-order modes. 
This filtering can be done by considering an expansion of the periodic density in Fourier modes, namely plane waves with wave vectors $\boldsymbol q^{(n)}$, and retaining a small set of \textit{modes}, i.e., sets $\{\boldsymbol q^{(n)} \}$ having the shortest lengths~\cite{skogvoll2021,Punke_2022}.
%, thus filtering out as well higher-order harmonics of the density field.  
In particular, one may write $\psi$ as
\begin{equation}
\psi(\boldsymbol{r},t) =\overline{\psi}(\boldsymbol{r},t) +\sum_{n=1}^N \eta_n(\boldsymbol{r},t) \rm{e}^{ \mathbbm {i} \boldsymbol q^{(n)}\cdot \boldsymbol r},
\label{eq:1modetri}
\end{equation}
with imaginary unit $\mathbbm{i}$, local average density $\overline{\psi}(\boldsymbol{r},t) =\langle \psi \rangle$ and amplitude functions $\eta_n(\boldsymbol{r},t)$ with respect to a set of reciprocal lattice vectors. For a triangular lattice, an one-mode approximation, $\psi_\text{1M}$, can be used with $N=6$, $\lbrace \boldsymbol q^{(n)} \rbrace_{n=1}^6 = q_0 \lbrace (0,\pm 1), (\pm \sqrt{3}/2, \pm 1/2) \rbrace$. 
For the square lattice, a two-mode approximation, $\psi_\text{2M}$, can be considered with, $N=8$, $\lbrace \boldsymbol q^{(n)} \rbrace_{n=1}^8 = q_0 \lbrace (0,\pm 1), (\pm 1,0),(\pm 1,\pm 1) \rbrace$.
The amplitudes in \eqref{eq:1modetri} can be obtained by demodulation, i.e., the selection of a specific Fourier mode
\begin{equation}
\begin{split}
\eta_n&= \rm{e}^{ -\mathbbm {i} \boldsymbol q^{(n)}\cdot r }\; \mathbb{F}^{-1}\left(\rm{e}^{-2\pi \boldsymbol a^2(\boldsymbol k-\boldsymbol q^{(n)})^2} \;\mathbb{F}(\psi)\right).
\end{split}
\label{eq:demodul}
\end{equation} 
    
The approximations via Fourier modes have to be defined with respect to a set of reference lattice vectors $\lbrace \boldsymbol q^{(n)} \rbrace_{n=1}^N$ describing the undeformed state, entering \eqref{eq:demodul}.
This reference inherently needs to be adapted to the specific settings with a lack of generality. 
For instance, a rotation of the reciprocal lattice vectors is necessary for rotated crystal structures. 
Here, we propose a simplified routine to filter out higher-order Fourier modes of the density field, which does not require the choice of a specific reference set of reciprocal-space vectors and remove part of the stress due to residual compression. This is achieved by defining the following smoothed Heaviside-function $\Theta$ in Fourier-space
$\chi(\boldsymbol{k})= 
e^{-2\pi \boldsymbol{a}^2 (\boldsymbol{k}^2-1) \Theta (\boldsymbol{k}^2-1)} $ 
for the triangular symmetry and 
$\chi(\boldsymbol{k})= 
e^{-2\pi \boldsymbol{a}^2 (\boldsymbol{k}^2-2) \Theta (\boldsymbol{k}^2-2)} $ 
for the square symmetry. 
This translates to considering contributions within a circle in the Fourier space, including the reciprocal space vectors entering the one- and two-mode approximation, respectively.
By exploiting $\chi$, we define the filtered density field  
    \begin{equation}\label{eq:filtered}
    \psi_\chi = \mathbb{F}^{-1}(\mathbb{F}(\psi) \chi),
    \end{equation}
and determine the stress as $\sigma_{ij}(\psi_\chi)$ from the equations \eqref{eq:stress_tri} and \eqref{eq:stress_square}. 
We compare the results obtained by computing the stress field from different density approximations in the following.
    
\section{Numerical Benchmarks}
 We compute numerical solutions of \eqref{eq:hpfc} with a Fourier pseudo-spectral method for discretization in space, enforcing periodic boundary conditions on an uniform grid with element size $\rm{d} x, \rm{d}y <1$, resulting in a resolution of $\approx 100$ grid points per unit cell. 
 The (inverse) Fourier transforms are performed with an algorithm based on the FFTW2 library. 
 For time discretization, we use a first-order IMEX time-stepping scheme for updating $\psi$  and a first-order fully implicit time-stepping scheme for updating $\boldsymbol{v}$. 
 A constant time step size of $\Delta t=0.1$ ensures numerical convergence~\cite{Punke_2023}.
We compare the stress fields \eqref{eq:stress_tri} and  \eqref{eq:stress_square}, obtained by the following density fields: i) \textit{unfiltered} density field $\psi$ as from numerical solutions of Eq.~\eqref{eq:hpfc}; ii) \textit{one-mode} approximation $\psi_\text{1M}$ (for triangular symmetry); (iii) \textit{two-mode} approximation $\psi_\text{2M}$ (for square symmetry); iv) \textit{filtered} density $\psi_\chi$ from Eq.~\eqref{eq:filtered}.
   % \begin{itemize}
   %     \item density field $\psi$ as from numerical solutions of Eq.~\eqref{eq:hpfc} (\textit{unfiltered})
   %     \item \textit{one-mode} approximation $\psi_\text{1M}$ (for triangular symmetry)
   %     \item \textit{two-mode} approximation $\psi_\text{2M}$ (for square symmetry)
   %     \item density $\psi_\chi$ from Eq.~\eqref{eq:filtered} (\textit{filtered})
   % \end{itemize}
%In the following, we refer to these cases as unfiltered, 1-Mode, 2-Mode, and filtered density.

\begin{figure}[h]
\centering
\includegraphics{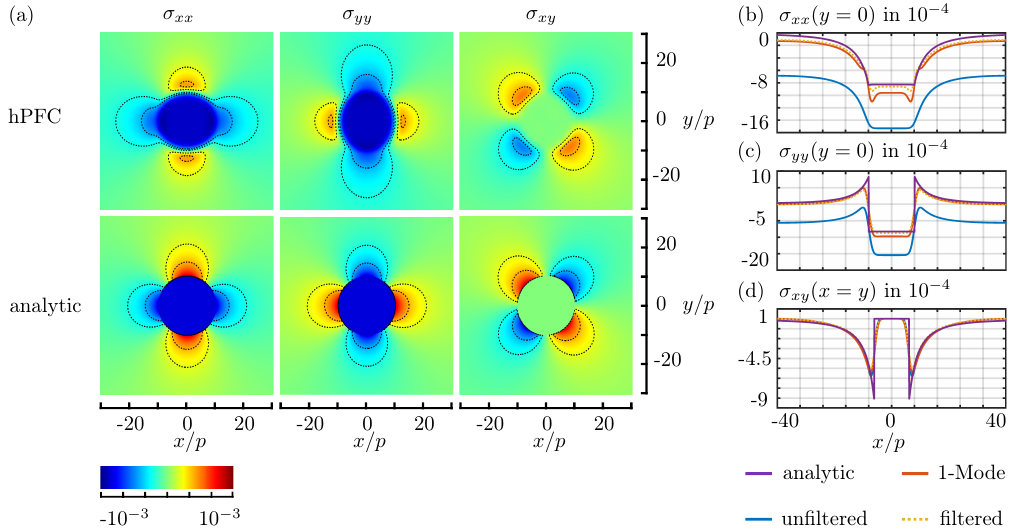}
\caption{Benchmark of the stress definition, Eq. \eqref{eq:stress_tri}, through the Eshelby inclusion problem. (a) Stress field components extracted from the result of an hPFC simulation with filtered density (first row) and the analytic solution for an inclusion in an infinite medium (second row). (b) Comparison of $\sigma_{xx}(x)$ for these solutions and the results obtained through the unfiltered density and the one-mode approximation. $y=0$ corresponds to the center of the inclusion. (c) Similar to (b) but for $\sigma_{yy}(x)$. (d) Similar to (b) but for $\sigma_{xy}(x)$ at $x=y$. Lengths are scaled with the atomic spacing along the $x$-axis, $p$.
	}
	\label{fig:eshelby}
    \end{figure}

We consider first the benchmark problem of an elastic inclusion, i.e., the so-called Eshelby problem~\cite{mura}. An eigenstrain formulation that enforces a spatially dependent lattice parameter through a quantity similar to $\beta$ has been proposed for the PFC framework~\cite{counterexamples,Punke_2022}. 
Here we consider the same setting within the hPFC model for a crystal with a triangular lattice. 
We set a spatially dependent, but constant in time, lattice parameter
 \begin{equation}
    \begin{split}
\beta = 1-\left( 1-\dfrac{1}{1+\varepsilon^*}\right) \xi,\qquad \qquad
\xi = \dfrac{1}{2}\left(1-\rm{tanh}\left(\dfrac{|\boldsymbol{r}|-R}{\varepsilon}\right)\right)
    \end{split}
    \label{eq:eshelby}
    \end{equation}
with $\varepsilon^{*}    = 0.01$, $R = 10p$, $\varepsilon = p$ and a domain $\Omega =[-50p,50p]\times [-25\sqrt{3}p,25\sqrt{3}p]$. 
We let the system relax until a steady state is found. 
The material and model parameters are set to $\overline{\psi}=0$, $\lambda=1.04$, $\kappa = 1$, $\delta=1$, $\rho=2^{-6}$, $\Gamma = 2$ \cite{skogvollhpfc}. 

\begin{figure}
    \centering
       \includegraphics{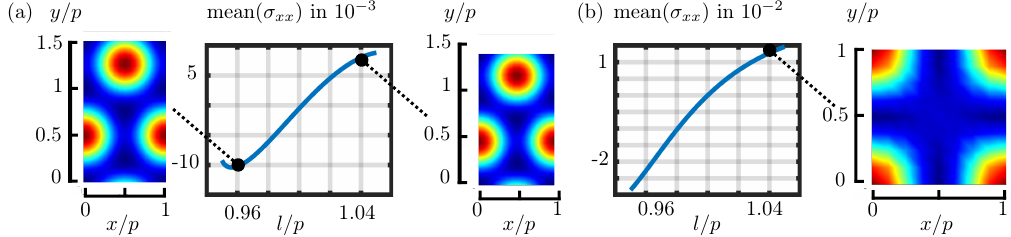}
    \caption{
     The residual stress in an one-unit PFC model with (a) triangular symmetry and (b) square symmetry as a function of simulation domain size.
    The insets exemplify the shape of the triangular PFC in the simulations.
    The different PFC models were initiated using the same model parameters for the problems considered in the text.
    The square PFC model for this set of parameters gives a residual stress ($-4.2\cdot 10^{-5}$) at $l=1p$ almost one order of magnitude smaller than the triangular PFC ($-5.6\cdot 10^{-4}$). Note that the latter case is in quantitative agreement with the offset shown in Fig.~\ref{fig:eshelby} between stress computed from the unfiltered and filtered density.
    }
    \label{fig:residual_stress_function_of_domain_size}
\end{figure}

Figure~\ref{fig:eshelby}(a) provides first a qualitative comparison of the mechanical stress components computed from the filtered density $\psi_\chi$ in the hPFC simulation through equation \eqref{eq:hpfc} and the analytic solution for an inclusion in an infinite medium\cite{mura,li}. 
A good agreement between the analytic solution and the hPFC simulation is obtained. 
A few differences may be ascribed to different assumptions underlying the derivation of the analytical solution and the simulations framework. 
The latter, in brief, accounts for a finite-size system with periodic boundary conditions, a smooth transition between inclusion and the matrix, as well as nonlinear elasticity effects. 
The former instead is derived for an infinite system, a sharp interface between the inclusion and the matrix, and in the regime of linear elasticity \cite{counterexamples}.
Figure~\ref{fig:eshelby}(b)--(d) compares in detail the stress fields for the different methods mentioned above, i.e., for different approximations of the density field. 
If no filter is applied, the stress field deviates by a constant residual pressure from the analytic solution, i.e., a deviation is observed for $\sigma_{xx}$ and $\sigma_{yy}$ which is not present for $\sigma_{xy}$. 
This residual pressure arises from the equilibrium periodicity deviating from $q_0=1$.
By closer inspection, we characterize this deviation in detail as reported in Fig.~\ref{fig:residual_stress_function_of_domain_size} (for both the triangular and square symmetry) and comment further below.
%See Fig.~\ref{fig:residual_stress_function_of_domain_size}.
The one-mode approximation and the filtered density $\psi_\chi$ in Fig.\ref{fig:eshelby} show both results approaching quite well the analytical solution. 
They both filter out higher harmonics in the density field by construction, enforcing the inspection of the main modes corresponding to periodicity $p$ and wavenumber $q_0$.  
Interestingly, slightly different behavior is observed close to the inclusion. The approximation provided by the stress field computed from the filtered density $\psi_\chi$ best matches the analytic stress values within the inclusion.

To verify that the difference between the stress computed from unfiltered and filtered density is due to residual pressure because of lattice/simulation domain incompatibility, we initiate a $1\times 1$ unit cell triangular PFC with the same parameters as given in the simulation of Fig.~\ref{fig:eshelby}. 
Then, we vary the simulation domain $[l,\sqrt{3}l]$ between $l\in [0.95p,1.05]$ in intervals of $0.001p$.
For each value, we initiate the PFC and let it evolve according to Eq. \ref{eq:pfc} for 10 000 time steps to reach complete equilibrium.
We then calculate the mean of the residual stress $\sigma_{xx}$ in the simulation domain. 
The result is shown in Fig. \ref{fig:residual_stress_function_of_domain_size}(a). It shows that, for a system size smaller than $p$, the model experiences compression (negative stress), while for system sizes larger than $p$, it experiences stretching tension (positive stress). 
Furthermore, at $l = 1.00p$, there is a slight (negative) residual stress which shows that the PFC equilibrium lattice unit is, in fact, slightly larger than $p$ for the given set of parameters. 
This residual stress disappears at approximately $l=1.002p$, showing that the actual unit lattice is $0.2\%$ larger than $p$. 
We perform the same analysis for the square lattice by adapting the unit cell and report the result in \ref{fig:residual_stress_function_of_domain_size}(b). 
In this case, the deviation is not as large as for the triangular PFC. 
This indicates that the equilibrium lattice unit of the square PFC model is closer to $p$ owing to a different differential operator encoding more details by enforcing two stable wavelengths (in $\mathcal{L}_1\mathcal{L}_2$) compared with the single wavelength encoded for the triangular symmetry (in $\mathcal{L}_1$).

\begin{figure}[h]
	\centering
	\includegraphics{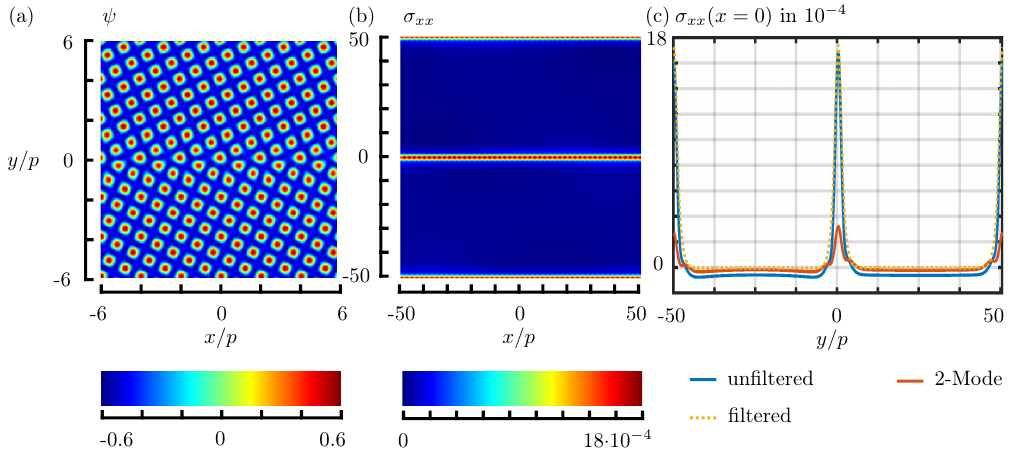}
	\caption{
 Stress fields along a symmetric tilt $\Sigma 5$ GB. (a) Density field $\psi$ (close up). (b) Stress field $\sigma_{xx}$  obtained through the filtered density $\psi_\chi$. (c) Comparison of $\sigma_{xx}(y)$ for different modifications of the density field at $x$ corresponding to the center of the domain.}	\label{fig:sigma5}
    \end{figure}

 As the second benchmark problem, we consider the simple but instructive case of a symmetric tilt $\Sigma 5$ grain boundary (GB), consisting of two square symmetric crystals with $\overline{\psi}=-0.3$ and a relative rotation angle of $53.1^\circ$. 
 We prepare the system with the following material parameters: $\lambda=0.233$, $\kappa = 1/3$, $\delta=0$, $\rho=2^{-6}$, $\Gamma = 2$. We take the domain size as a multiple of the (nominal) periodicity of the rotated lattices~\cite{mellenthin,blixt}, resulting in $\Omega \approx 100 \text{UC}$.
 Atomistic methods like molecular dynamics simulations~\cite{han} show that along the GB, localized positive stress $\sigma_{xx}$ occurs, while in the bulk crystals (away from the GB), the stress fields vanish rapidly.
 Figure~\ref{fig:sigma5} shows a close-up of the simulated setup as well as the resulting stress fields $\sigma_{xx}$ for an uncorrected density field, a two-mode approximation (reciprocal lattice vectors have to be rotated by $\pm 53.1/2^\circ$ for demodulating according to \eqref{eq:demodul}) 
 and the filtered density field. 
 All the methods provide good qualitative agreement with the expected stress behavior. The decay of this field is found to follow an exponential law \cite{hirth}. The unfiltered density shows a very small --but not zero-- residual stress, consistent with the above results (c.f. Fig.~ \ref{fig:residual_stress_function_of_domain_size}). To compute the two-mode approximations, two reference systems must be considered, namely oriented as the top and bottom bulk crystals in Fig.~\ref{fig:sigma5}(a), with an arbitrary interpolation at the GB. Thus, such a description cannot accurately describe GBs, which may be crucial for complex morphologies and dynamics \cite{han}. Using the filtered density $\psi_\chi$ allows for the evaluation of the stress field of the GB approaching the results with unfiltered density and filtering out residual isotropic pressure contribution. Therefore, this approach can be exploited to inspect specific properties of these extended defects. We remark that for computing $\psi_\chi$, neither the local rotation angles nor the exact position of the GB has to be known.

\begin{figure}[h]
	\centering
	\includegraphics{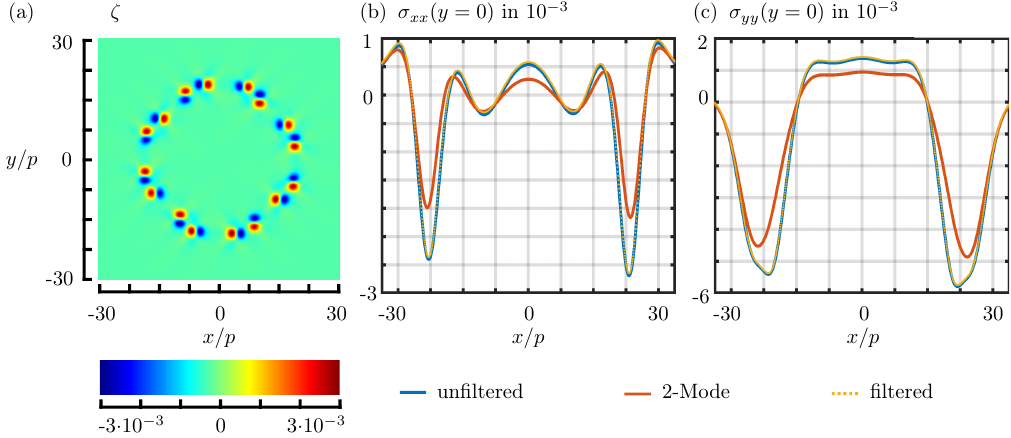}
	\caption{Stress fields of a circular rotated inclusion inside a relaxed bulk crystal (close up).
(a) Incompatibility field $\zeta$ to visualize the defects around the inclusion. (b) Comparison of $\sigma_{xx}(x)$ for different modifications of the density field at $y=0$ corresponding to the center of inclusion. (c) Comparison for $\sigma_{yy}(x)$.}
	\label{fig:inclusion}
    \end{figure}

As a third benchmark problem, we simulate a spherical inclusion in a square lattice rotated by a relatively small angle with respect to a surrounding relaxed bulk matrix, resulting in an arrangement of dislocations at the inclusion-matrix interface, which evolves until the annihilation of all defects (see, e.g., Refs.~\cite{heinonen2016,salvalaglio2019npj}). 
The spherical inclusion considered here has a radius of $20p$ and is rotated by $5^\circ$ with respect to the (unrotated) bulk crystal. 
The simulation domain is $\Omega = 500 \text{UC}$ with material and model parameters as in Fig.~\ref{fig:sigma5}. 
Defects around the spherical inclusion form upon relaxation of the initial condition, cf. Fig.~\ref{fig:inclusion} (a) (the incompatibility field $\zeta =\left( \epsilon_{ik} \epsilon_{jl} \sigma_{kl,ij} -\kappa \sigma_{kk,ll}\right)/2\mu$  with shear modulus $\mu \approx 0.055$ and Levi-Civita-symbol $\epsilon_{kl}$ is shown to visualize the defects~\cite{skogvoll2020}). In Fig.~\ref{fig:inclusion}~(b)--(c), the stress $\sigma_{xx}(x,0)$ and $\sigma_{yy}(x,0)$ are shown. In this case, the stress values for the unfiltered and filtered densities match well with very small residual values in the matrix for the former case ($\sigma_{xx}\approx 5\cdot 10^{-5}$), similarly to the case of the $\Sigma 5$ GB in the square lattice (see Fig.~\ref{fig:sigma5}).
However, the stress field obtained from the two-mode approximation of the density is significantly smoothened. This may lead to a loss of details at defects and affect dynamic aspects in out-of-equilibrium settings. Also in this case, the better candidate for inspecting defect-induced deformation is identified in the stress field computed from the filtered density field.
%leading to significantly smaller values of $\sigma_{xx}\approx 1.8\cdot 10^{-8}$.
%Similarly to the previous cases, the stress field computed from the filtered density field retains details similar to the uncorrected case while filtering out spurious effects in bulk regions. 
%The same qualitative behavior holds for $\sigma_{yy}$, cf. Fig.~~\ref{fig:inclusion}(c). 

\section{Conclusion}
In this study, we focused on evaluating the elastic field in phase-field crystal (PFC) models. 
In particular, we considered the hPFC model, an advanced formulation that includes elastic relaxation timescales based on classical physical principles and, thus, an accurate description of lattice deformations.

We considered different approaches to derive the elastic field from the microscopic density field, i.e., the PFC order parameter, relying on coarse-graining procedures.
A simple and novel numerical routine providing filtering of higher harmonics in the density field is proposed as a suitable approach to inspecting deformations induced by inclusion and defects. With the aid of this routine, we also characterize the emergence of residual stresses in the widely-used classical PFC model based on the Swift-Hohenberg energy functional. This is more relevant for the formulation encoding a single lengthscale (here shown for the triangular lattice) than approaches enforcing two lengthscales (here shown for the square lattice). We envisage that methods enforcing periodicity with a higher degree of detail, i.e., involving more lengthscales or definitions of correlation as proposed in Ref.~\cite{greenwood}, are more suitable for removing residual stress effects.

Finally, we remark that the method proposed here can be readily applied to the microscopic density field computed by other governing equations, i.e., it can be generally exploited within PFC models. Also, the simple settings for numerical simulations discussed in this work may be used as benchmarks for testing both numerical implementation and elastic field evaluation for current and novel PFC formulations.

\begin{acknowledgement}
We acknowledge fruitful discussions with Luiza Angheluta. MP and MS acknowledge support from the German Research Foundation (DFG) under Grant No.~SA4032/2-1. Computing resources have been provided by the Center for Information Services and High-Performance Computing (ZIH) at TU Dresden and by J\"ulich Supercomputing Center under grant PFAMDIS.
\end{acknowledgement}

\vspace{\baselineskip}
%% The style of the following references should be used in all documents.


\begin{thebibliography}{1}

\bibitem{Elder2002}
K.~R. Elder, M. Katakowski, M. Haataja, and M. Grant.
\newblock {\em Phys. Rev. Lett.} \textbf{88}, 245701 (2002).

\bibitem{Elder2004}
K.~R. Elder and M. Grant.
\newblock {\em Phys. Rev. E} \textbf{70}, 051605 (2004).

\bibitem{Emmerich2012}
H. Emmerich, H. L{\"{o}}wen, R. Wittkowski, T. Gruhn, G.~I.
  T{\'{o}}th, G. Tegze, and L. Gr{\'{a}}n{\'{a}}sy.
\newblock {\em Adv. Phys.} \textbf{61}, 665 (2012).

\bibitem{Punke_2022}
M. Punke, S.~M. Wise, A. Voigt, and M. Salvalaglio.
\newblock {\em Model. Simul. Mater. Sci. Eng.} \textbf{30}, 074004 (2022).

\bibitem{counterexamples}
M. Salvalaglio, K. Chockalingam, A. Voigt and W. Dörfler. 
\newblock {\em Examples Counterexamples} \textbf{37}, 100067 (2022).


\bibitem{Stefanovic2006}
Stefanovic, Peter and Haataja, Mikko and Provatas, Nikolas.
\newblock {\em  Phys. Rev. Lett.} \textbf{96(22)}, 225504 (2006).

\bibitem{heinonen2014}
V. Heinonen, C.~V. Achim, K.~R. Elder, S. Buyukdagli and T. Ala-Nissil\"a.
\newblock {\em  Phys. Rev. E} \textbf{89(3)}, 032411 (2014).

\bibitem{heinonen2016}
V. Heinonen, C.~V. Achim, J. M. Kosterlitz, S.-C. Ying, J. Lowengrub  and T. Ala-Nissila. 
\newblock {\em  Phys. Rev. Lett.} \textbf{116.2}, 024303 (2016).


\bibitem{skaugen2018}
A. Skaugen, L. Angheluta and J. Viñals.
\newblock {\em Phys. Rev. Lett.} \textbf{121}, 255501 (2022).

\bibitem{skogvoll2021}
V. Skogvoll, A. Skaugen and L. Angheluta. 
\newblock {\em Phys. Rev. B} \textbf{103}, 224107 (2021).

\bibitem{skogvollhpfc}
V. Skogvoll, M. Salvalaglio and L. Angheluta.
\newblock {\em Model. Simul. Mater. Sci. Eng.} \textbf{30}, 084002 (2022).


\bibitem{Burns_2022}
D. Burns, N. Provatas and M. Grant.
\newblock {\em  Model. Sim. Mater. Sci. Eng.} \textbf{30(6)}, 064001 (2022).


\bibitem{salvalaglio2019npj}
M. Salvalaglio, A. Voigt and K.~R. Elder.
\newblock {\em  npj Comput. Mater.} \textbf{5}, 48 (2019).

\bibitem{salvalaglio2020}
M. Salvalaglio, L. Angheluta, Z.-F. Huang, A. Voigt, K.~R. Elder and J. Viñals.
\newblock {\em J. Mech. Phys. Solids} \textbf{133}, 103856 (2020).


\bibitem{Punke_2023}
M. Punke, S.~M. Wise, A. Voigt, and M. Salvalaglio.
\newblock {\em PAMM} \textbf{23},  e202200112 (2023).

\bibitem{mura}
T. Mura.
\newblock {\em Micromechanics of Defects in Solids (Berlin: Springer)} (1982).


\bibitem{li}
S. Li, R. Sauer and G. Wang. 
\newblock {\em Acta Mech.} \textbf{179}, 67–90 (2005).

\bibitem{mellenthin}
J. Mellenthin, A. Karma and M. Plapp.
\newblock {\em Phys. Rev. B} \textbf{78.18}, 184110 (2008).


\bibitem{blixt}
K.~H. Blixt and H. Hallberg.
\newblock {\em  Model. Simul. Mater. Sci. Eng.} \textbf{30.1}, 014002 (2021).


\bibitem{han}
J. Han, S.~L. Thomas and D.~J.~Srolovitz.
\newblock {\em  Prog. Mater. Sci.} \textbf{98}, 386-476 (2018).

\bibitem{skogvoll2020}
V. Skogvoll, L. Angheluta, A. Skaugen and J. Viñals.
\newblock {\em Phys. Rev. B} \textbf{103}, 014107 (2020).

\bibitem{hirth}
P.~M. Anderson, J.~P. Hirth and J. Lothe.
\newblock {\em Theory of Dislocations (Cambridge University Press)} (2017).

\bibitem{greenwood}
M. Greenwood, N. Provatas, and J. Rottler, {\em Phys. Rev. Lett}. \textbf{105}, 045702 (2010).

%\bibitem{chen}
%N. Chen,  L.-L. Niu, Y. Zhang, X. Shu, H.-B. Zhou, S. Jin, G. Ran, G.-H. Lu and F. Gao. 
%\newblock {\em  Sci. Rep.} \textbf{6.1}, 36955 (2016).

%\bibitem{warrenNumeric}
%M. Cheng and J.~A. Warren.
%\newblock{\em J. Comput. Phys.} \textbf{227}, 12, 6241-6248 (2008).


%\bibitem{baskaranNumeric}
%S.~M. Wise, C. Wang and J.~S. Lowengrub.
%\newblock {\em SIAM J. Numer. Math.} \textbf{47}, 3, 2269-2288 (2009).
%\bibitem{baskaranNumeric}
%A. Baskaran, J.~S. Lowengrub, C. Wang and S.~M. Wise.
%\newblock {\em SINUM} \textbf{51}, 5, 2851-2873 (2013).

%\bibitem{dongNumeric}
%H. Gomez and X. Nogueira.
%\newblock {\em Comput. Methods Appl. Mech. Eng.} \textbf{249}, 52-61 (2012).

%\bibitem{ChengNumeric}
%K. Cheng, C. Wang and S.~M. Wise.
%\newblock {\em Commun. Comput. Phys.} \textbf{26}, 1335-1364 (2019).

%\bibitem{parkNumeric}
%J.-H. Park, A. Salgado and S.~M. Wise.
%\newblock arXiv:2204.07247 (2022).

%\bibitem{cox2002exponential}
%S.~M. Cox and P.~C. Matthews.
%\newblock {\em J. Comput. Phys.} \textbf{176}, 430 (2002).

%\bibitem{elsey2013simple}
%M. Elsey and B. Wirth.
%\newblock {\em ESAIM Math. Model. Numer. Anal.} \textbf{47}, 1413 (2013).

%\bibitem{wang2021thermodynamically}
%C. Wang and S.~M. Wise.
%\newblock {\em J. Math. Study} \textbf{55}, (2022).





%\bibitem{goldenfeld}
%N. Goldenfeld, B.~P. Athreya and J.~A. Dantzig. 
%\newblock {\em  Phys. Rev. E} \textbf{72}, 020601 (2005).

%\bibitem{athreya}
%B.~P. Athreya, G. Nigel and J.~A. Dantzig. 
%\newblock {\em  Phys. Rev. E} \textbf{74}, 011601 (2006).

%\bibitem{salvalaglio2022msme}
%M. Salvalaglio and K. R. Elder.
%\newblock {\em Model. Simul. Mater. Sci. Eng.} \textbf{30}, 053001 (2022).

%\bibitem{eshelby57}
%J. D. Eshelby.
%\newblock {\em  Proc. R. Soc. A} \textbf{241}, 376–96 (1957).


%\bibitem{eshelby59}
%J. D. Eshelby.
%\newblock {\em  Proc. R. Soc. A} \textbf{252}, 561–9 (1959).



%Han, Jian, Spencer L. Thomas, and David J. Srolovitz. "Grain-boundary kinetics: A unified approach." Progress in Materials Science 98 (2018): 386-476


%\bibitem{salvalaglio2018}
%M. Salvalaglio, R. Backofen, K.~R. Elder and A. Voigt.
%\newblock {\em  Phys. Rev. Materials} \textbf{2(5)}, 053804 (2018).



\end{thebibliography}
\end{document}